%% file: semantic-knowledge-graph.tex
\documentclass[10pt, conference, compsocconf]{IEEEtran}

\usepackage{wrapfig}
\usepackage{algpseudocode}
\usepackage{array}
\usepackage{listings}
\usepackage{comment}
\usepackage{cite}
\usepackage[numbers,sort&compress]{natbib}
\usepackage{graphicx}
\usepackage{multirow}
\usepackage{float}
\usepackage{chngpage}
\usepackage{amsthm}
\usepackage{amsmath}
\usepackage{amssymb}
\usepackage{array}
\usepackage[T1]{fontenc}
\usepackage{tabularx}
\usepackage{enumerate}
\usepackage{nameref}
\usepackage{textcomp}

\begin{document}
%
% paper title
% can use linebreaks \\ within to get better formatting as desired
\title{The Semantic Knowledge Graph:  A compact, auto-generated model for real-time traversal and ranking of any relationship within a domain}
\author{\IEEEauthorblockN{Trey Grainger, Khalifeh AlJadda, Mohammed Korayem, and Andries Smith}

\IEEEauthorblockA{ CareerBuilder, Norcross, GA, USA }

Email: \{trey.grainger, khalifeh.aljadda, mohammed.korayem, andries.smith\}@careerbuilder.com}
\maketitle

% make the title area

\input{abstract}

% IEEEtran.cls defaults to using nonbold math in the Abstract.
% This preserves the distinction between vectors and scalars. However,
% if the conference you are submitting to favors bold math in the abstract,
% then you can use LaTeX's standard command \boldmath at the very start
% of the abstract to achieve this. Many IEEE journals/conferences frown on
% math in the abstract anyway.

% no keywords

% For peer review papers, you can put extra information on the cover
% page as needed:
% \ifCLASSOPTIONpeerreview
% \begin{center} \bfseries EDICS Category: 3-BBND \end{center}
% \fi
%
% For peerreview papers, this IEEEtran command inserts a page break and
% creates the second title. It will be ignored for other modes.
\newcommand{\xhdr}[1]{{\vspace{6pt}\noindent\textbf{\textit{#1}}}}

\newenvironment{packed_itemize}{
	\begin{itemize}
		\setlength{\itemsep}{0pt}
		\setlength{\parskip}{0pt}
		\setlength{\parsep}{02pt}
		\vspace{-2pt}
	}{\end{itemize}}

\vspace{12pt}  
\input{intro}
\input{related_work}
\input{methodology}

\input{system_implementation}
% no \IEEEPARstart
\input{experiments_results}
\input{future_work}

\input{conclusion}
\input{acknowledgment}
 \setlength{\bibsep}{0ex}
\vspace{-5pt}
\small{
	%\footnotesize
\bibliographystyle{abbrv}	
\bibliography{references}
}
% that's all folks
\end{document}

%% file: abstract.tex
\begin{abstract}
This paper describes a new kind of knowledge representation and mining system which we are calling the Semantic Knowledge Graph. At its heart, the Semantic Knowledge Graph leverages an inverted index, along with a complementary uninverted index, to represent nodes (terms) and edges (the documents within intersecting postings lists for multiple terms/nodes). This provides a layer of indirection between each pair of nodes and their corresponding edge, enabling edges to materialize dynamically from underlying corpus statistics. As a result, any combination of nodes can have edges to any other nodes materialize and be scored to reveal latent relationships between the nodes. This provides numerous benefits: the knowledge graph can be built automatically from a real-world corpus of data, new nodes - along with their combined edges - can be instantly materialized from any arbitrary combination of preexisting nodes (using set operations), and a full model of the semantic relationships between all entities within a domain can be represented and dynamically traversed using a highly compact representation of the graph. Such a system has widespread applications in areas as diverse as knowledge modeling and reasoning, natural language processing, anomaly detection, data cleansing, semantic search, analytics, data classification, root cause analysis, and recommendations systems. The main contribution of this paper is the introduction of a novel system - the Semantic Knowledge Graph - which is able to dynamically discover and score interesting relationships between any arbitrary combination of entities (words, phrases, or extracted concepts) through dynamically materializing nodes and edges from a compact graphical representation built automatically from a corpus of data representative of a knowledge domain. The source code for our Semantic Knowledge Graph implementation is being published along with this paper to facilitate further research and extensions of this work.\footnote[1]{http://github.com/careerbuilder/semantic-knowledge-graph/tree/dsaa2016}
\end{abstract}

%% file: intro.tex
\vspace{-18pt}
\section{Introduction}
Graphs are a well-studied class of data structures used to model relationships (edges) between entities (nodes).  Knowledge bases in general, and ontologies specifically, model a domain by defining how different entities within the domain are related. Such knowledge bases are most commonly represented as a graph, and both the nodes and the edge relationships between nodes in that graph must be explicitly modeled either manually by a domain expert or automatically leveraging an ontology learning system. Because building such a knowledge base typically requires explicitly modeling nodes and edges into a graph ahead of time, this unfortunately presents several limitations to the use of such a knowledge graph:

\begin{packed_itemize}
	
	\item Entities not modeled explicitly as nodes have no known relationships to any other entities.
	\item Edges exist between nodes, but not between arbitrary combinations of nodes, and therefore such a graph is not ideal for representing nuanced meanings of an entity when appearing within different contexts, as is common within natural language.
	\item Substantial meaning is encoded in the linguistic representation of the domain that is lost when the underlying textual representation is not preserved: phrases, interaction of concepts through actions (i.e. verbs), positional ordering of entities and the phrases containing those entities, variations in spelling and other representations of entities, the use of adjectives to modify entities to represent more complex concepts, and aggregate frequencies of occurrence for different representations of entities relative to other representations.
	\item It can be an arduous process to create robust ontologies, map a domain into a graph representing those ontologies, and ensure the generated graph is compact, accurate, comprehensive, and kept up to date.
\end{packed_itemize}

We propose a new system for modeling relationships between entities that overcomes these limitations. This system, which we refer to as a Semantic Knowledge Graph, is aimed at extracting and representing the knowledge of a domain automatically from a corpus of documents representative of that domain. The underlying representation ultimately encodes the semantic relationships between words, phrases, and extracted concepts in such a way that those relationships can later surface to expose new insights about the interrelationships between all entities within the domain.

This kind of system has numerous applications which we will explore. It can be used to automatically discover sets of related terms within a domain, to represent and disambiguate multiple meanings of the same phrases, to power semantic search by dynamically expanding user queries to conceptually-related keywords/phrases, to identify trending topics across time-series data, to build a content-based recommendation engine, to perform data cleansing on lists by scoring how relevant each items is to the list, to perform document summarization by detecting the importance of each phrase and entity within a document, and to do predictive analytics on time series data.

In its most basic use case (a corpus of free-text documents), a Semantic Knowledge Graph can be leveraged to automatically discover domain-specific relationships between entities within a domain. Given a corpus of documents also containing some amount of structured information (specific fields for titles, categories, dates, or other specific kinds of entities), it will treat each of those field types as a new edge that can be traversed between any two nodes co-occurring within the same documents with some (specifiable) minimum frequency.

One of the novelties of the system is that a layer of indirection exists between each node and the edge connecting it to any other node. Instead of explicitly defining an edge connecting two nodes with a predetermined relationship, as most graph databases are designed, a Semantic Knowledge Graph instead materializes edges during traversal between any two nodes based upon the intersection of the document sets to which both of the nodes link. Furthermore, because the edges between nodes are dynamically materialized based upon the set of shared documents to which they both link, this means that it is also possible to dynamically materialize new nodes by combining existing nodes (through their underlying sets of documents) in any arbitrarily-complex way. This subsequently means that any arbitrarily-complex nodes (for example, any linguistic combination of character sequences, terms, and term sequences) can also be decomposed into their minimum constituent parts (terms related by position within documents) when building the graph, enabling a highly-compressed graph representation which is capable of reconstituting and traversing every existing relationship within a knowledge domain. 

As a result of maintaining all of the corpus occurrence statistics about each node, a Semantic Knowledge graph can also dynamically discover and score interesting relationships between any nodes based upon the statistical similarity of the nodes in any given context. The Semantic Knowledge Graph represents a novel new graph model which is both auto-generated and yet able to represent, traverse, and score every relationship represented within a corpus of documents representing a knowledge domain.

%% file: related_work.tex
\vspace{-10pt}
\section{Related Work}
Ontologies can be defined as explicit formal specifications of the terms within a domain and the relations among them \cite{gruber2009ontology}. Ontologies have become common across various domains for building vocabulary to be shared and used by domain experts. Many advantages can be gained by building a common vocabulary, including improving the re-usability of domain knowledge, enabling a common understanding of the structure of information, and providing the ability to analyze domain knowledge. 
Ontologies can be classified into three different categories~\cite{biemann2005ontology}: formal ontologies that have axioms and definitions in logic, terminological ontologies (e.g, WordNet~\cite{miller1995wordnet}), and prototype-based ontologies having typical instances or prototypes instead of axioms. 
Recently, large-scale knowledge bases that utilize ontologies (FreeBase~\cite{bollacker2008freebase}, DBpedia ~\cite{lehmann2015dbpedia}, and YAGO ~\cite{hoffart2013yago2,suchanek2007yago}) have been constructed  using structured sources such as Wikipedia infoboxes.  Other approaches (DeepDive~\cite{niu2012deepdive}, Nell2RDF ~\cite{zimmermann2013nell2rdf}, and PROSPERA~\cite{nakashole2011scalable}) crawl the web and use machine learning and natural language processing to build web-scale knowledge graphs. 

Existing work on ontologies and knowledge bases still suffers from significant limitations.  Manually-created knowledge bases are expensive and labor-intensive to build and maintain and are thus generally incomplete and have a tendency to grow out of date over time. While ontology learning systems are typically able to automate much of the ontology building (and sometimes maintenance) process, this comes at the expense of a loss of accuracy due to the replacement of human experts with more error-prone algorithms ~\cite{noy2000algorithm,daly2015development,krizhanovsky2013approach,lee2007automated,dong2014knowledge}.

Current ontology learning systems also throw away a substantial amount of information encoded within the textual content they are processing. For example, any entities not discovered as nodes during the ontology mining process have no known relationships to any other entities, regardless of whether those relationships were actually represented within the analyzed content (and just overlooked) during the ontology mining process. Furthermore, since most terms and phrases can take on alternate, nuanced meanings within different contexts, these nuanced meanings are often lost when representing terms and phrases as single nodes independent of the context in which they are used. Finally, a substantial amount of meaning is encoded in the nuanced linguistic representations present in a corpus of free-text content (terms, character sequences, term ordering, placement of words within phrases and paragraphs, and so on). Existing ontology creation approaches fail to adequately support the representation and scoring of relationships between these nuanced and complex interrelationships.

Our work improves upon current ontology mining approaches by creating a knowledge graph which can fully represent the nuanced relationships between every entity (term, phrase, or other textual representation) represented within a corpus of free-text documents, as well as traverse and score the strength of those relationships or of any combination of those relationships.

%% file: methodology.tex
\vspace{-10pt}
\section{Methodology}
\input{problem_description}
\input{model_structure}

%% file: problem_description.tex
\subsection{Problem Description}

Technology platforms are becoming increasingly more capable every day of interpreting and responding to domain-specific and personalized questions. Search engines and recommendation engines, in particular, can barely compete unless they leverage models containing deep insights into the kinds of questions being asked and - more importantly - the kinds of answers being sought. One of the most common ways of representing a domain in order to surface these insights is through the use of ontologies - combinations of taxonomies containing known entities, their properties, and their interrelationships. These ontologies can then be integrated into a search application in order to improve its ability to meet the end-user's information need. For example, if someone searches for the term \textit{server} in the information technology domain, it has a very different meaning (a computer server) than in the restaurant domain (a waiter/waitress), and if someone is using a job search engine, it could actually represent either meaning depending upon the user's context. Ontologies can help represent the relationships between entities such that they can be used to improve the accuracy of the system at meeting its users' information needs.

Ontologies are usually built manually by human experts, making them expensive to build, maintain, and update. To combat this, ontology learning systems, which attempt to automatically learn relationships from a domain and then map them into an ontology, are becoming more prevalent~\cite{navigli2004learning}. 

We would like to create a system that is able to automatically generate a graph representation of a knowledge domain simply from ingesting a corpus of data representing the domain, while simultaneously preserving all of the linguistic and statistical relationships between the keywords, phrases, and extracted entities from the corpus. Once a model (a graph) is built from this data, we can then leverage it to better understand the interrelationships between those words, phrases, and entities. 

Natural language, as represented in full-text documents, contains tremendous meaning compressed within its linguistic structures, represented through multiple levels of abstraction:
\begin{packed_itemize}
	\item \textit{Corpus}: a list of \textit{documents} representative of a knowledge domain	
	\item \textit{Document}: a list of \textit{fields} relating to each other through some underlying entity
	\item \textit{Field}: a grouping of zero or more \textit{term sequences} representative of a relationship with a document.
	\item \textit{Term Sequence}: an ordered representation of one or more \textit{terms}
	\item \textit{Term}: a \textit{character sequence} representing a known meaning (for example, a recognizable word)
	\item \textit{Character Sequence}: an ordered combination of one or more \textit{characters}
	\item \textit{Character}: a letter or symbol used within natural language (represents no meaning by itself)
\end{packed_itemize}

\begin{figure}
	
	\includegraphics [scale=0.34] {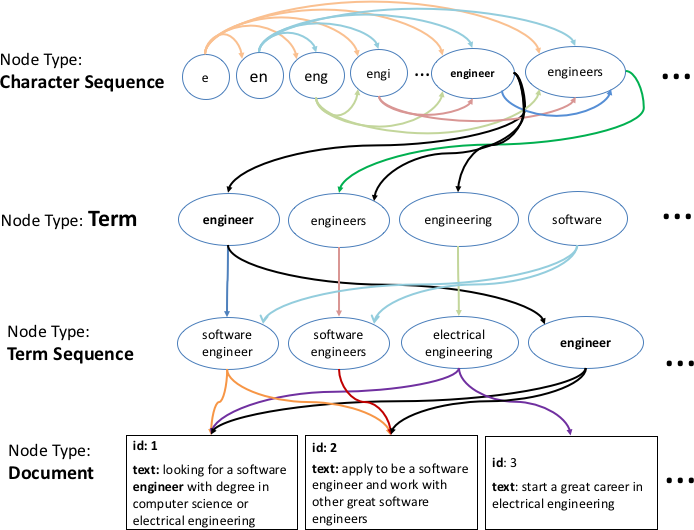}
	\caption{\textbf{Semantic relationships encoded in free text content}. \textit{Terms} are composed of one or more \textit{character sequences}, \textit{term sequences} are composed of one or more \textit{terms}, and \textit{documents} are composed of one or more \textit{fields} containing zero or more \textit{term sequences}.
		}
	\label{fig:nodes}
	\vspace{-5pt}
\end{figure}

While a corpus, document, and field are common concepts within the field of information retrieval, concrete examples of term sequences, terms, character sequences, and characters are presented in Figure~\ref{fig:nodes} for further explication. In this figure, you can see that the term sequence \textit{software engineering} is composed of the ordered sequence of the two terms \textit{software} and \textit{engineer}, and that the word \textit{engineering} contains the character sequence \textit{engineer}, which contains the characters \textit{e}, \textit{en}, \textit{eng}, ... \textit{engineer}, etc.

Our goal is to automatically generate a knowledge graph from an underlying corpus of documents. In order to avoid the previously mentioned pitfalls with manually generated ontologies and ontology learning systems, we need a way to fully preserve these nuanced semantic interrelationships embedded within a corpus of textual documents.

Our overarching goal is not simply to link entities with known relationships, however, but to actually present the ability to discover any arbitrarily-complex relationship between entities within the domain. Consider a typical search engine, where a user can query any keywords, phrases, or arbitrarily-complex combinations of character sequences, terms, or term sequences. We would like to be able to traverse our automatically-generated knowledge graph and instantly understand the nuanced meaning(s) represented by these arbitrarily-complex natural language queries.

To understand the significance of this goal, let's consider the way in which the meaning of terms is modified given their context. The term \textit{engineer} has a well-known abstract meaning, but when found inside the phase \textit{software engineer}, it takes on a much more limited interpretation. Similarly, the word \textit{driver} takes on two entirely different meanings when found near terms relating to computers (a \textit{hardware driver}) versus in contexts related to transporting goods (\textit{truck driver} or \textit{delivery driver}). While we tend to think of most terms and phrases as having a limited number of meanings, it is far more accurate to think of them as having a slightly different meaning in every unique context in which they are found. While terms and phrases usually share strong similarities in their intended meanings across contexts, by allowing both those strong similarities, as well as nuanced differences to surface during node traversals, we are able to discover the most important interrelationships between entities in any given context and thus much better represent the intended knowledge domain. Our Semantic Knowledge Graph model provides a compact representation of an entire knowledge domain (as represented within a corpus of documents) which accomplishes these goals.

%% file: model_structure.tex
\vspace{-5pt}
\subsection{Model Structure}\label{sec:modelStructure}

Consider an undirected graph $G=(V,E)$ where $V$ and $E\subset V\times V$
denote the sets of nodes and edges, respectively. We define the following: 
\begin{packed_itemize}
\item $ D = \{d_1,d_2,... ,d_m\}$ is a set of documents that represent a corpus that the Semantic Knowledge Graph will utilize to extract and score semantic relationships.
\item $ X = \{x_1,x_2, ..., x_k\}$ is a set of all items stored in $D$. These items could be keywords, phrases, or any arbitrary linguistic representation found within $D$.
\item $ d_i= \{x|x \in X\} $ where we can think of each document $d\in D$ as a set of items.  
\item $ T = \{t_1,t_2, ... t_n\}$ where $t_i$ is a tag which assigns an entity type to an item such as keyword, title, location, company, school, person, etc.
\end{packed_itemize}  
Given the previous notations, the set of nodes $V$ in our graph can be defined as $V=\{v_1,v_2, ..,v_n\}$  where $v_i$ stores an item $x_i \in X$ tagged with tag $t_j \in T$. While $D_{v_i} = \{d|x_i\in d, d\in D\}$ is a set of documents that contains item $x_i$ with its appropriate tag $t_j$. Finally, we define $e_{ij}$ as an edge between $(v_i,v_j)$ with a function $f(e_{ij}) =\{d\in D_{v_i}\cap D_{v_j}\}$ that stores on each edge the set of documents that contain both items $x_i$ and $x_j$ with their tags. On the other hand, we define $g(e_{ij},v_k) = \{d: d\in f(e_{ij})\cap D_{v_k}\}$ that stores on the edge $e_{jk}$ the common set of documents between $f(e_{ij})$ and $D_k$.

\vspace{-5pt}
\subsection{Materialization of Nodes and Edges}

Core to the SKG model is the idea that a layer of indirection exists between any two nodes $v_i$ and $v_j$ and the edge $e_{ij}$ that connects them. Specifically, instead of nodes being directly connected to each other through explicit edges, nodes are instead connected bidirectionally to documents, such that the edge $e_{ij}$ between node $v_i$ and $v_j$ is said to \textit{materialize} whenever $|f(e_{ij})| > 0$.

\begin{figure*}
	\begin{tabular}{@{}c@{}|c@{}|c@{}}
		\includegraphics [scale=0.28] {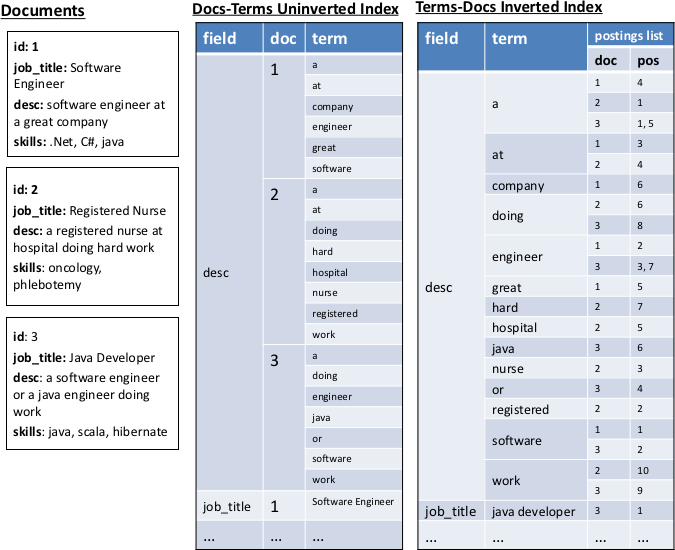}  		\hspace{0.2cm} &
		\hspace{0.2cm}
		\includegraphics [scale=0.28] {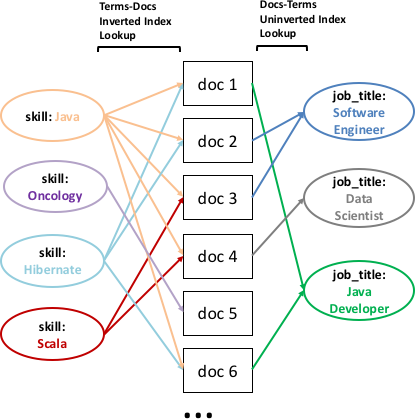} \hspace{0.2cm} &
		\hspace{0.3cm}
		\includegraphics [scale=0.28] {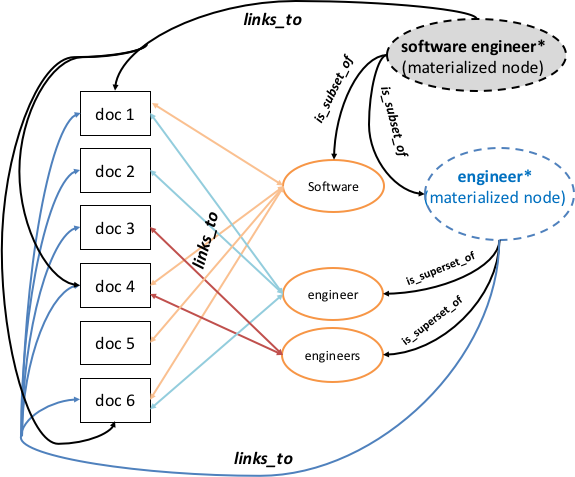} 
		\\
		(a) & (b)  & (c) 
	\end{tabular}
	\caption{ 
		\textbf{(a):} \textbf{Two indexes needed to power the Semantic Knowledge Graph}. The SKG leverages two indexes per field, a docs-terms uninverted index mapping documents to the terms they contain, and a terms-docs inverted index mapping every term in the corpus to a postings list of documents containing the term and the positions of the term within each document.
		\textbf{(b):} \textbf{Materialization of edges using shared documents}. Only terms which share documents have an edge, and the weight of the edge will be later calculated based upon the statistical distribution of shared documents. \textbf{(c):} \textbf{Materialization of new nodes through shared documents}. New nodes can be dynamically formed (materialized) through specifying any arbitrary combination of character sequences, terms, and term sequences, finding the underlying document set they all match, and leveraging this document set for subsequent traversals.}
	
	\label{fig:index_materialization_edges_materialization_newnodes}
\end{figure*}

Thus, in order to traverse the graph from source node $v_i$ to destination node $v_j$, our system requires a lookup index linking node $v_i$ to a set of documents, as well as a separate lookup index which can map from those documents to node $v_j$ or other nodes to which a traversal may need to occur. We refer to this first index as our terms-docs inverted index, and to the second as our docs-terms uninverted index, both shown in Figure \ref{fig:index_materialization_edges_materialization_newnodes} (a). These two indexes enable us to model all terms as nodes within the graph and to materialize and traverse from any node to any other node through the sets of shared documents between the nodes, as shown in Figure \ref{fig:index_materialization_edges_materialization_newnodes} (b).

Because edges materialize during graph traversal based upon an intersection of documents to which both nodes are connected, this means that we can form an edge between \textit{any} entity that is representable by an underlying set of documents to which it is linked. Thus, instead of being restricted to only using predefined entities from our terms-docs index, it is also possible to dynamically materialize new nodes on the fly based upon any combination of terms, as shown in Figure ~\ref{fig:index_materialization_edges_materialization_newnodes} (c). 

Since complex representations of entities can be materialized as nodes from arbitrary combinations of existing terms, this enables us to also decompose complex entities into individual terms (with positional relationships) for persistence in the underlying terms-docs inverted index. Through this process of decomposing our corpus into individual terms, the documents in which the terms appear, and the positions in those documents where the terms appear, we can thus create a highly compressed and lossless representation of every relationship within our original corpus. Then, at traversal time, we can materialize nodes representing any representation found within the original corpus, as well as edges connecting any materialized or predefined nodes to other nodes.

\vspace{-5pt}
\subsection{Scoring Semantic Relationships}

The Semantic Knowledge Graph (SKG) is able to score and represent the strength of the semantic relationship between entities on the edge connecting them. For example, if we don't know how semantically related the keyword \textit{java} is to the keyword \textit{hadoop}, we can utilize the SKG to score the relationship between these two terms. 
To score a semantic relationship between item $x_i$ and item $x_j$ using the SKG, we materialize source node $v_i$ (holding the documents linked to by $x_i$) and destination node $v_j$ (representing the set of documents containing $x_j$). 

The simple use case for scoring semantic relationships is to score directly connected nodes $v_i$ and $v_j$. In this case we query the terms-docs inverted index for item $x_i$ tagged with $t_j$, and as a result we get back $D_{vi}$. Then we query the terms-docs inverted index again for $x_j$ tagged with $t_k$ to get $D_{vj}$. An edge $e_{ij}$ will be created between $v_i$ and $v_j$ if $f(e_{ij}) \ne \phi$. We call the $D_{vi}$ our $foreground$ document set $D_{FG}$, while $D_{BG} \subseteq D$ is our $background$ document set. The hypothesis behind our scoring technique is that if $x_i$ tends to be semantically related to $x_j$, then the presence of $x_j$ in the $foreground$ document set $D_{FG}$ should be above the average presence of $x_j$ in $D_{BG}$. We utilize The $z$ score to evaluate this hypothesis:
\[
z(v_i,v_j) = \frac{y-n*p}{\sqrt{n*p(1-p)}}
\] 
Where $n=|D_{FG}|$ is the number of documents in our $foreground$ document set, $p=\frac{|D_{v_j}|}{|D_{BG}|}$ is the probability of finding the term $x_j$ with tag $t_k$ in the $background$ document set, and $y = |f(e_{ij})|$ is the number of documents containing both $x_i$ and $x_j$.
 
In many cases, we will want to traverse multiple levels of depth $n > 2$ to find and score relationships between more than just two nodes. For example, we may traverse from the entity $java$ to $big\ data$ to $hadoop$, such that the weight assigned to the edge between $big\ data$ and $hadoop$ would be more meaningful if it were also conditioned upon the the path it took to arrive at $big\ data$ through $java$.  Our system accomplishes this across $n$ nodes and a path $P={v_1,v_2,..,v_n}$, where each node stores an item $x_i$ with its tag $t_j$. To apply the same $z(v_i,v_j)$ between nodes, but conditioning this score based upon the entire path $P$, the only changes are

\vspace{-7pt}
\[ D_{FG} =
\begin{cases}
f(e_{ij})				& \quad \text{if } n = 3\\
\{\bigcap\limits_{i=1,j=i+1,k=j+1}^{n-3}g(e_{ij},D_{v_k})\}       & \quad \text{if } n > 3\\
\end{cases}
\]
\vspace{-6pt}

 while $y = |D_{FG}\cap D_{v_n}|$. We normalize the $z$ score using a sigmoid function to bring the scores in the range $[-1,1]$. We call the normalized score the \textit{relatedness score} between nodes where $1$ means completely positively related (likely to always appear together), while $0$ means no relatedness (just as likely as anything else to appear together), and -1 means completely negatively related (unlikely to appear together).

While the relatedness score provides a weight on each edge corresponding to the strength of the semantic relationship between two nodes, since this score is calculated at traversal time, it is also possible to substitute in other scoring functions depending upon the use case at hand. Popularity (total count of overlapping documents) is another function that may be appropriate for simpler use cases, for example.

\vspace{-5pt}
\subsection{Discovering Semantic Relationships}

\begin{figure*}
	\hspace{-0.25in}
\begin{tabular}{@{}c@{}|c@{}|c@{}}
\includegraphics [scale=0.28] {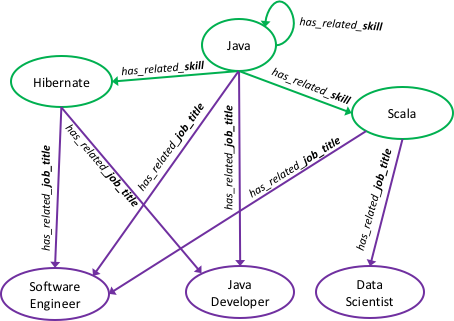} \hspace{0.2cm}&
\hspace{0.1cm}
	\includegraphics [scale=0.28] {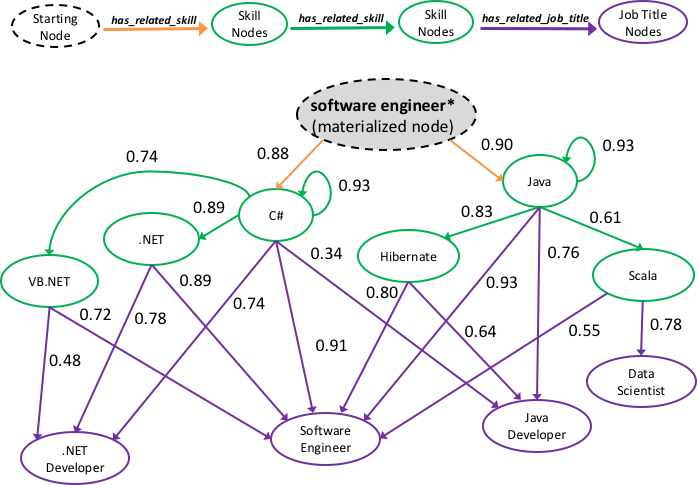} 
\hspace{0.2cm}	
	&
		\includegraphics [scale=0.28] {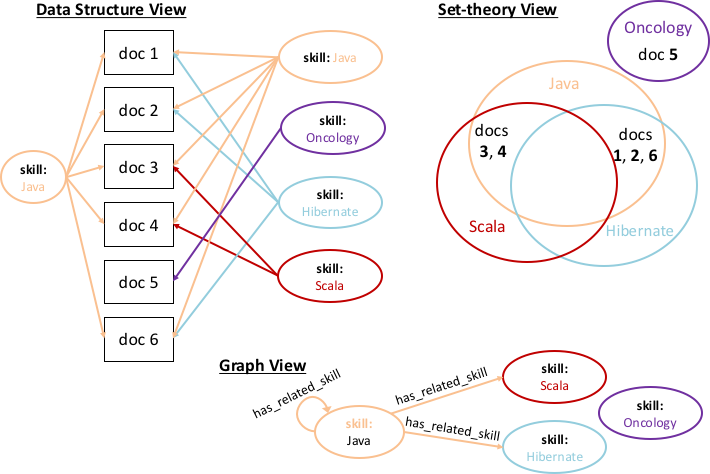} \\
(a) & (b)  & (c) 
\end{tabular}
\caption{ (a): \textbf{Graph representation of node traversal}. Models a graph traversal across edges of a specific relationship type (tag). The first traversal is from the starting node of \textit{Java} to each node for which it holds a \textit{has\_related\_skill} edge. The second traversal is to each subsequent node for which a \textit{has\_related\_job\_title} edge is found. (b): \textbf{Multilevel graph traversal}. This example traverses from a materialized node, through all \textit{has\_related\_skill} edges, then from that level of nodes again through each of their \textit{has\_related\_skill} edges, and finally from those nodes to each of their \textit{has\_related\_job\_title} edges. The weights are calculated using the entire traversed path in this example, though it is also possible to consider calculate weights independent of the path using only each pair of directly connected nodes.	(c): \textbf{Three representations of a traversal}. The Data Structure View represents the underlying links from term to document to term in our underlying data structures, the Set Theory view shows the relationships between each term once the underlying links have been resolved, and the Graph View shows the abstract graph representation in the semantics exposed when interacting with the SKG.
	}
\label{fig:graph_representaion_multilevel_graph_traversal_three_view_traversal}
\end{figure*}

The SKG is very powerful at surfacing hidden relationships between nodes in real time. Furthermore, this model enables materialization of nodes and extraction of relationships using those materialized nodes. In order to discover related items with a specific tag $t_k$ to an item $x_i$ with tag $t_j$, we start by querying the inverted index for the item $x_i$, which we assign as node $v_i$ corresponding with document set $D_{v_i}$. We query the docs-terms uninverted index for the tag $t_k$ and we store the retrieved documents as $D_{t_k} = \{d|x \in d, x:t_k\}$. We define $V_{v_i,t_k} = \{v_j|x_j\in d, d\in D_{t_k}\cap D_{v_i}\}$ where ${v_j}$  is  a node that stores an item $x_j$, and we define $V_{v_i,t_k}$ as the set of nodes that stores items with potential relationship with $x_i$ of type $t_k$ (See Figure \ref{fig:graph_representaion_multilevel_graph_traversal_three_view_traversal} (a)). Finally, we apply $\forall v_j \in  V_{v_i,t_k}, relatedness_{(v_i,v_j)}$ to score the relationship between $v_i$ and $v_j$, which enables us to rank those relationships and pick the top $m$ relationships or define a threshold $t$ to accept only relationships with $relatedness(v_i,v_j)>t$. This operation of relationship discovery can occur recursively, as shown in Figure \ref{fig:graph_representaion_multilevel_graph_traversal_three_view_traversal} (b), to discover and drill into multiple levels of relationships.

%% file: system_implementation.tex
\vspace{-5pt}
\section{System Implementation} \label{system_implementation}

Our implementation of the Semantic Knowledge Graph leverages the Apache Lucene/Solr search engine for many of its needed data structures. The data structures leveraged include the underlying inverted index that is used to find preexisting nodes, the document set intersection logic necessary to materialize new nodes and edges from the terms-docs inverted index, and the docs-terms uninverted index that is necessary to traverse across the edges materialized between nodes.

Since Apache Solr serves as a web server, we leverage it as a framework to expose a RESTful API around our SKG implementation.

In order to build the knowledge graph, one simply needs to send a corpus of documents to the Semantic Knowledge Graph API. These documents will contain one or more fields, typically with at least one field contain raw text, and optionally with one or more additional fields containing some more structured information about the document. For the use case of employment search, for example, we could use the command in Table~\ref{tab:API} to add some job postings to the SKG.

\begin{table}
\begin{tabular}{|m{8cm}|}
\hline
\begin{lstlisting}[columns=fullflexible,breaklines=true,upquote=true]
curl -H 'Content-type:application/json'
http://localhost:8983/solr/semantic-knowledge-graph/update  -d 
'[{ "id" : "job1", 
     "title" : "Data Scientist", 
     "skills": ["machine learning","spark"], 
     "keywords": "Seeking a senior-level data scientist with experience with spark and machine learning..."},
  { "id" : "job2", 
     "title" : "Registered Nurse", 
     "skills": ["er","trauma", "phlebotomy"], 
     "keywords": "Come join the top-rated hospital in the region..."}
]'
\end{lstlisting} \\
\hline
\end{tabular}
	\vspace{5pt}
\caption {Adding Documents to the SKG}
	\vspace{-20pt}
\label{tab:API}
\end{table}

The most important thing to note here is that documents are added to the graph, but no explicit relationships between entities need to be modeled. Instead, the SKG will later allow us to discover relationships between entities - in this case job titles, skills, and keywords - through statistical analysis of how those entities are found together or absent across the entire corpus of documents. Figure \ref{fig:graph_representaion_multilevel_graph_traversal_three_view_traversal} (c) visually demonstrates how the underlying data structure and the intersection of sets of documents work together to form a traversable graph model.

Once an entire corpus of documents has been loaded into the SKG, we can now issue queries to the system to traverse and score the relationships between entities. Table~\ref{tab:sampleQuery} shows an example query and response from the graph.

\begin{table}
		\vspace{-15pt}
	\begin{tabular}{|m{3.2cm}|m{5.2cm} |}
		
		\hline
		Request & Response \\ 
		\hline
		\begin{lstlisting}[columns=fullflexible,escapechar=\%]
		
{ "starting_node": [ 
   "keywords:\"data science\""
   ],
  "nodes": [
   { "type": "job_title",
     "limit": 1,
     "discover_values": true,
     "nodes": [
      { "type": "skills",
        "limit": 3,
        "discover_values": true,
        "values": [ "java"]
        }]}]}

          %%
		\end{lstlisting}
		
		& \begin{lstlisting} [columns=fullflexible,breaklines=true]
{ "nodes": [{
    "type": "job_title",
    "values": [{
      "name": "Data Scientist",
      "relatedness": 0.989,
      "popularity": 86.0,
      "fg_popularity": 86.0,
      "background_popularity": 142.0,
      "nodes": [{
        "type": "skills",
        "values": [
          { "name": "Machine Learning",
            "relatedness": 0.97286,
            "popularity": 54.0,
            "foreground_popularity": 54.0,
            "background_popularity": 356.0 }, 
          { "name": "Predictive Modeling",
            "relatedness": 0.94565,
            "popularity": 27.0,
            "foreground_popularity": 27.0,
            "background_popularity": 384.0 },
          { "name": "Artificial Neural Networks",
             "relatedness": 0.94416,
             "popularity": 10.0,
             "foreground_popularity": 10.0,
             "background_popularity": 57.0 },
          { "name": "Java",
             "relatedness": 0.76606,
             "popularity": 37.0,
             "foreground_popularity": 37.0,
             "background_popularity": 17442.0
           }]}]}]}]}
			\end{lstlisting} \\
			\hline
			
	\end{tabular}
	\caption{Sample Graph Traversal Request}
	\label{tab:sampleQuery}
			\vspace{-20pt}
\end{table}
This request asks the system to find the top job title associated with the phrase \textit{data science}, and then to find up to three skills, including \textit{java}, sorted by how similar they are to the job title \textit{data science} (the previous node in the traversal).

By making our implementation of the SKG model a plugin for Apache Solr, we were able to leverage a pre-built inverted index, an uninverted index, as well as a rich set of text analysis libraries (tokenizers and token filters) to model documents. This allowed us to focus on the graph semantics, document set intersections, scoring models, and graph traversal API required to implement the SKG without needing to reimplement most of the already well-studied information retrieval structures upon which the SKG relies. Instead of re-implementing a query parser, this also allowed us to make full use of existing tools to map character sequences, terms, and term sequences into their underlying document set representations.

Since an inverted index can be implemented using multiple underlying data structures, this also allows us to easily leverage highly efficient and compressed data structures, such as Lucene's Finite State Automata/Transducers, for more efficient compression and traversal of nodes within the SKG.~\cite{Daciuk2011}

%% file: experiments_results.tex
\vspace{-5pt}
\section{Experiments and Results}
While the SKG is a generally applicable model to any domain representable by documents with overlapping references to the same entities, we focused our testing on use cases within the job search domain, leveraging datasets provided to us by CareerBuilder, one of the largest job boards in the world. 

For our experiments, we leveraged two datasets: 1) a collection of 3 million job postings, and 2) a collection of 1 million job seeker resumes containing a total of 3 million employment history sections (representing prior jobs held by a given job seeker). While these two datasets could have been combined into a single graph, we only had the need to use a single dataset at a time and therefore maintained each dataset in a separate SKG for the following experiments. All of our experiments leveraged the SKG implementation described in the \nameref{system_implementation} section, which has been open sourced along with the publication of this paper.

In terms of performance, the SKG was able to easily traverse through and gather millions of nodes in just a few milliseconds (on commodity servers) in our experiments when the relatedness score was not needed. For most of these same queries tested utilizing the relatedness score, the SKG request completed in tens to hundreds of milliseconds, though some very intensive queries traversing multiple levels of nested-relationships were observed to take several seconds to complete. Observed times are thus quite fast considering the dynamic nature of the node and edge materialization, making the SKG suitable for integration in real-time search and natural language processing systems.
\vspace{-8pt}
\input{data_cleansing}
\vspace{-8pt}
\input{predictive_analytics}

\vspace{-8pt}
\input{document_summarization}

%% file: data_cleansing.tex
\subsection{ Data Cleansing} 
Data Scientists spend a considerable amount of their time - 60\% according to a 2016 survey - cleaning and organizing data sets~\cite{CrowdFlowerDataScience}. Most datasets contain some dirty data, particularly when free text content is involved. While we have previously described how the SKG is able to discover relationships embedded within a corpus of documents, it is just as good at ranking user-supplied relationships.

\textit{Use Case:} As an example use case, we leveraged the SKG to clean a list of relationships mined from search engine query logs using a similar methodology to that described in~\cite{Jargon,7004213}. The idea here is that users who conduct similar searches often search for related terms and phrases. For example, someone who searches for \textit{registered nurse} will often also search for \textit{RN}, \textit{nurse}, \textit{ER}, \textit{hospital} and so on. Someone who searches for \textit{java} will often also search for \textit{software engineer}, \textit{java developer}, and so on. After obtaining a list of search terms mapped to their co-occurring terms, we decided to use the SKG to find the weight of the edge between each pair of co-occurring terms.

Since the relatedness score is normalized between -1 (perfectly negatively related), 0 (no relationship), and 1 (perfectly positively related), we use 0.5 as our threshold between whether something is likely to have a strong relationship (0.5 to 1.0) or likely to have a weak relationship (0 to 0.5).

\textit{Experiment Setup:} To setup our experiment, we indexed 3 million job postings into an SKG. We then used this SKG to traverse 2.26 million co-term pairs, traversing across the shared \textit{has\_related\_term} edge between the nodes for each term. For each traversal, we analyzed the weight of the edge (the relatedness score), and we added all term pairs with a relatedness score below 0.5 to a blacklist. The end result was the blacklisting of 78\% of the co-term pairs (1.77 million blacklisted). Table ~\ref{tab:res-datacleansing} shows some examples of co-term pairs which were kept and which were blacklisted.

\begin{table}[]
	\centering
	\caption{Samples for the co-terms  cleaned by SKG}
	\label{my-label}
	
	\begin{tabular}{lll}
		\hline
		Term                          &  Co-term              & Blacklisted?    \\ \hline
		system support           &  it manager      & Yes       \\
		senior buyer       &  customer service manager& Yes       \\
		leasing consultant            & manufacturing manager & Yes    \\ 
		programmer & engineering manager & Yes  \\
		product requirement documents & sows                & No         \\
		events                        & wedding coordinator & No       \\ 
		electrical engineering & cad designer & No \\	 
		\hline	
	\end{tabular}
	\label{tab:res-datacleansing}
		\vspace{-15pt}
\end{table}

\textit{Results}: From the blacklist generated from the SKG, we asked an independent data analyst to randomly select 500 of these blacklisted pairs and tag them as either \textit{related} or \textit{not related}. The threshold for something being related, in this case, is whether the data analyst believed a user would wish to see the co-term suggested in a search experience when performing a search for the original term. As a result of his analysis, the Data Analyst determined that 25 of the 500 blacklisted terms were actually \textit{related}, while 475 were correctly identified by the SKG as \textit{not related}. The final results thus showed that the SKG removed 78\% of the terms while maintaining a 95\% accuracy at removing the correct noisy pairs from the input data.

%% file: predictive_analytics.tex
\subsection{Predictive Analytics:} The SKG is also effective at performing predictive analytics in order to estimate future behavior based on analysis of past behavior. 

Concretely, given a set $x_1 = \{{ x_{1_1} ... x_{1_m}}\}$ through $x_n$ of feature vectors describing a state at time $t_1$ through $t_n$, we would like to predict likely subsequent states at times $t_n +1 ... t_m$. 

With a modified scoring function, the SKG materializes nodes which can be interpreted as either consequent or antecedent of an association rule, with edge scores corresponding to the confidence of these rules~\cite{hipp2000algorithms}.

\begin{figure}
	
	\includegraphics [scale=0.20] {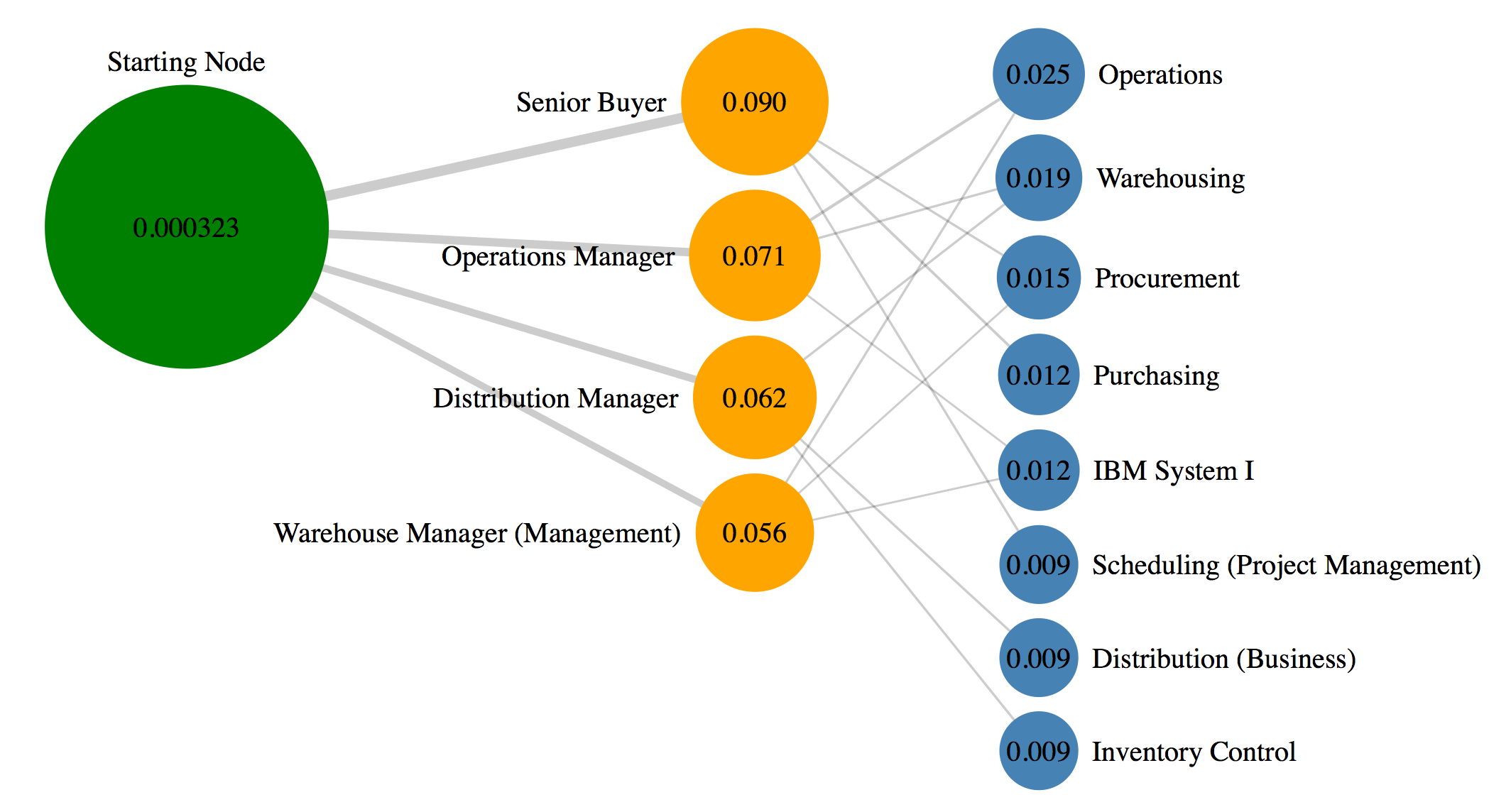}
	\vspace{-10pt}
	\caption{\textbf{Predictive analytics (consequent scoring)}. Assume a jobseeker has a job title of \textit{Logistics Manager}, the skill of \textit{Distribution (Business)}, and additionally some experience with the keyword \textit{purchasing}.  The figure shows this starting materialized node with its support on the left. The figure highlights the results for the top five predicted job titles with the middle circles, with the highest confidence job title being \textit{Senior Buyer} with a confidence of 0.09. The top skills are predicted jointly with the job title in the circles on the right, with \textit{Operations} as the highest confidence skill, with confidence of 0.025.
	}
	\vspace{-10pt}
	\label{fig:career_path}
	\vspace{-5pt}
\end{figure}

\textit{Career Pathing Use Case:} We used the SKG against resume data to characterize and predict employment histories. The tag types indexed include features such as extracted skills and keywords, normalized job titles, job level, location, and duration (in months) of employment. The SKG generates estimates characterizing a hypothetical next job in terms of combinations of node types indexed. For example, we can generate the maximal confidence consequent of the next job title and the skill most likely to be used at the next job. The SKG implementation also allows support and confidence thresholding through a normalized \textit{min\_count} parameter. Additionally, with the ability to materialize edges and nodes using query parameters, the SKG allows for much more fluid construction of antecedents and consequents. 

\textit{Experiment Setup:} We tested the SKG's predictive capabilities using resumes from one million job seekers. We parsed and extracted the tag types of location, job level, job title, skills, and keywords for the most recent three employment history entries of each resume. For each tag set (corresponding to an employment history), we index each tag type appended with an index indicating recency. We additionally use \textit{consequent} scoring, allowing edge scores to be interpreted as the confidence of association rules, with the new scoring function $c(v_i, v_j) = \frac{| D_{FG} \bigcap D_{v_j} |}{ | D_{FG} |}$. Prediction proceeds by materializing a node encoding the predictor features (which must be less recent than the most recent tag set), then traversing the graph through the \textit{next\_most\_recent\_t} edge for an arbitrary tag $t$.

\textit{Results:} Using data on what career paths thousands of other job seekers have taken, we can answer the question "Given my current position and skills, what are my next most likely positions?'' Figure \ref{fig:career_path} demonstrates an example answer.

	The most direct application of this predictive capability is in providing recommendations for job seekers looking to take the next step in their careers. As of the time of writing, such a system was still in a research phase. Our approach relies on the dynamic edge materialization of the SKG to discover viable job titles, which are then filtered by compensation (and in the future, experience) constraints to ensure recommended jobs represent a step forward. 
 
\textit{Search Expansion Use Case:} Recruiter search expansion represents another application of the SKG's predictive capabilities. A common problem when recruiting for high-demand jobs is a scarcity of applicants. Searching for applicants who match the skills and title of an in-demand job may be too restrictive, but recruiters don't always have the domain knowledge to expand their search for fitting candidates. A semantic search engine represents an adequate solution to this problem, but without the concept of career progression, semantic search ignores a pool of trainable candidates. Given an original candidate search, $q_0$, we consider the problem of expanding the query while retaining the highest possible probability that the additional candidates represent a "good fit''. One measure of trainability is the probability that the candidate would advance to match $q_0$ in their next job independent of the recruiter. By this definition, our problem can be reduced to finding a maximal confidence antecedent of a given starting node. 
 
\textit{Experiment Setup:} We tested the search expansion capabilities of the SKG using the same career path corpus containing one million resume examples. We modify our scoring function again, to an \textit{antecedent} scoring function, which evaluates the confidence of a rule defined  $V_k \rightarrow v_1$, where $v_1$ is the starting node and $V_k$  the set of nodes traversed up to the index $k$ (excluding $v_1$). Given a path $P = v_1, v_2, ... v_i$:
%\vspace{-2pt}
\[
	\vspace{-5pt}
a(v_i, v_k) =
\begin{cases}
\frac{| D_{v_k} \bigcap D_{FG} |}{| D_{BG} |} & \quad \text{if $v_i$ is a starting node} \\
\\
\frac{| D_{v_k} \bigcap D_{FG} |}{| \bigcap\limits_{j = 2}^{j = i} ( D_{v_j} ) \bigcap D_{BG} |}   & \quad \text{ otherwise } \\
\end{cases}
	\vspace{-3pt}
\]
\vspace{1pt}

In order to isolate the effect of career progression we modified our materialized starting node by explicitly excluding examples that matched the query in earlier employment history entries.  We then traverse the graph along the \textit{has\_less\_recent\_t} edge for arbitrary tag type $t$. 

\begin{figure}
 	
 	\includegraphics [scale=0.20] {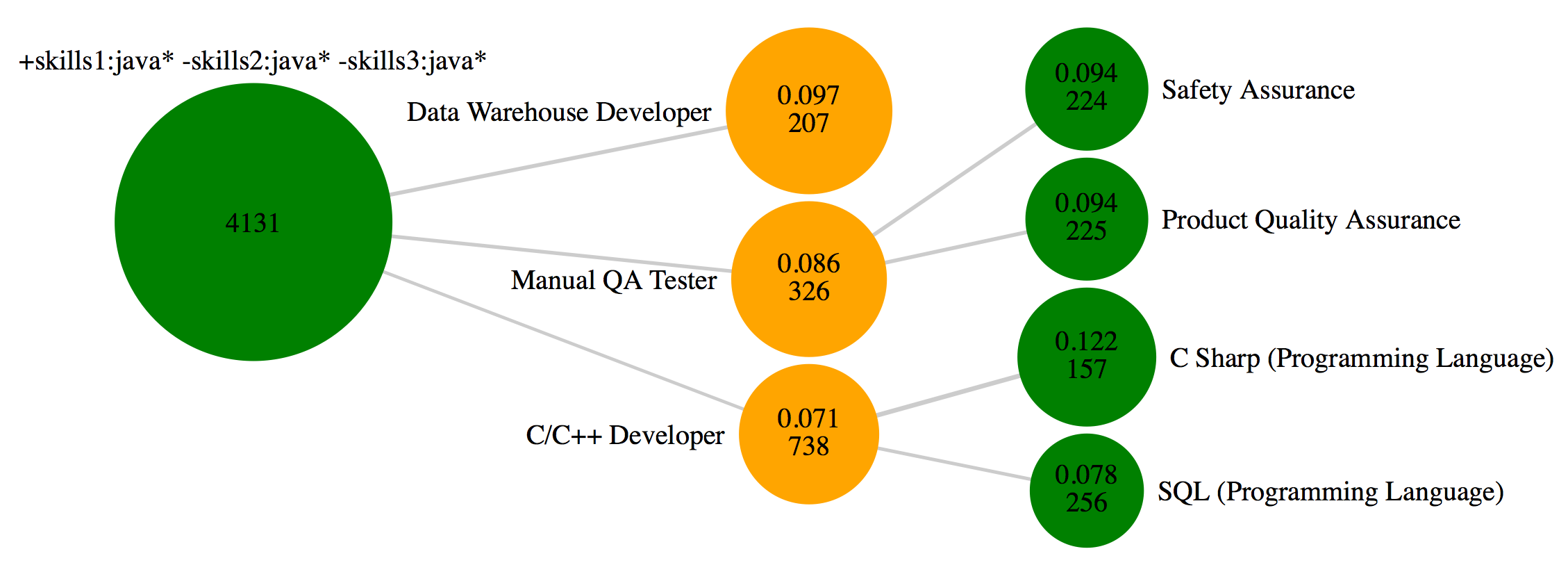}
 		\vspace{-10pt}
 	\caption{\textbf{Predictive analytics (antecedent scoring)}. An example of query expansion for a $q_0$ of \textit{skills:Java*}. The nodes joined by edges on the middle and right form the combined antecedent, with the query result set forming the consequent. The top number on the rightmost nodes equals the confidence of the combined antecedent $\rightarrow$ starting node rule, while the top number for the middle column represents the confidence of the single-item antecedent $\rightarrow$ starting node rule. Correspondingly, the bottom number indicates the support (times one million) for each rule. }
 		\vspace{-10pt}
 	\label{fig:search_expansion}
 		\vspace{-5pt}
 \end{figure}

\textit{Results:}  Figure \ref{fig:search_expansion} shows the results for an example query. Note the relatively high confidences for the distantly related job titles and skills, which are unlikely to be returned by a semantic search engine. Although applications for this use case are still in development, our approach would be to use the SKG to generate expansions, which can then be selectively applied based on confidence and support thresholds. 
 
%% file: document_summarization.tex
\subsection{Document Summarization}
Another interesting application of the SKG is the identification of the most important topics within a document. In any given text document, some words are going to be highly-related to the topic of the document, while others will be unimportant. With the SKG, we can score every entity within a document to determine its significance to the topic of the document. This takes us from a full text document to a much more compressed summarization of the document including only its most important components.

\begin{table}[t]
	\begin{adjustwidth}{-0.0in}{-0.0in}
		
	\begin{tabular}{|m{5.9cm}|m{2.8cm} |}	
		\hline
		Request & Response \\ 
		\hline
		\begin{lstlisting}[columns=fullflexible,breaklines=true]
Job Title: Big Data Engineer

REQUIREMENTS:
Bachelor's degree in Computer Science or related discipline...
2+ years of hands-on implementation experience(preferably lead engineer) working with a combination of the following technologies: Hadoop, Map Reduce, Pig, Hive, Impala, ...

IDEAL ADDITIONAL EXPERIENCE: 
Strong knowledge of noSQL of at-least one noSQL database like HBase and Cassandra.
3+ years' programming/scripting languages Java and Scala, python, R, Pig
2+ years' experience with spring framework
Experience in developing the full life-cycle of a Hadoop solution. This includes creating the requirements analysis, design of the technical architecture, design of the application design and development, testing, and deployment of the proposed solution...
Understanding of Machine Learning skills (like Mahout)
Experience with Visualization Tools such as Tableau
...
		\end{lstlisting}
		& 		
\begin{tabular}{ll} %{|m{0.5cm}|m{0.5cm} }

\multicolumn{1}{l}{} data engineer & 0.96 \\
\multicolumn{1}{l}{}         hive & 0.82 \\
\multicolumn{1}{l}{}         pig & 0.82 \\
\multicolumn{1}{l}{}         hadoop & 0.8 \\
\multicolumn{1}{l}{}         mapreduce & 0.79 \\
\multicolumn{1}{l}{}         nosql &  0.71 \\
\multicolumn{1}{l}{}         hbase & 0.66 \\
 \multicolumn{1}{l}{}        impala & 0.6 \\ 
 \multicolumn{1}{l}{}        python & 0.56 \\    
 \multicolumn{1}{l}{}        cassandra & 0.56 \\      
 \multicolumn{1}{l}{}        scala & 0.56 \\
 \multicolumn{1}{l}{}machine learning & 0.49 \\ 
 \multicolumn{1}{l}{}        tableau & 0.39 \\
 \multicolumn{1}{l}{}        mahout & 0.37 \\
 \multicolumn{1}{l}{}        analytics & 0.36 \\
 \multicolumn{1}{l}{}        java & 0.36 \\

\end{tabular}	
   \\ \hline		
	\end{tabular}
	\vspace{5pt}
	\caption{Document Summarization}
	\label{tab:sampleDoc}
		\end{adjustwidth}
\vspace{-10pt}
\end{table}

\textit{Experiment Setup}: We indexed 3 million job posting documents into the SKG implementation discussed in the System Implementation section. 

Our goal was to then take a new document not already represented in the graph and to have the graph score how related each of the entities found within the new document is to the document itself. While we could have used the individual keywords within the document as our starting nodes, we instead employed an entity extractor on the document as a preprocessing step. The purpose of leveraging the entity extractor was so that we could work with phrases as our nodes (e.g. \textit{senior software engineer} and \textit{registered nurse}) as opposed to only single keywords (e.g. \textit{senior}, \textit{software}, \textit{engineer}, \textit{registered}, and \textit{nurse}). 

Our next step was to specify a foreground query (which yields the set of documents $D_{FG}$ from our model), which in this case should represent the topic of our document. Because our corpus was composed of job posting documents, which have job titles and descriptions, we are able to simply leverage the job title of the document as our foreground query. In other scenarios where no category for the document is known a priori, it is possible to instead leverage other statistics from the terms-docs inverted index, such as tf-idf scoring of each term within the document, to find the set of most globally interesting terms within the document~\cite{ramos2003using}. This list of terms can then be used to materialize a foreground query that combines the top most globally interesting terms found within the document.

Once we generate our foreground query (the topic of the document), we then send each of the phrases from the document to the SKG, asking the graph to score how relevant they are to the topic of the document.

Table \ref{tab:sampleDoc} demonstrates an example document run through the SKG. While parts of the document were omitted for the sake of space, you can see that the top scoring nodes returned from the SKG are an excellent summarization representing the most important entities within the document. If someone wanted a quick overview about what this job posting is about, reviewing this ranked list of phrases would provide a very condensed summary. Furthermore, one could request additional traversals from this summary list and find the most related other nodes which were not actually present in the original documents.

Such document summarization has many applications. The summarized list of nodes can be sent to an information retrieval engine to build a document-based query, which creates a form of content-based recommendation engine. You could use the weights of each nodes to highlight the important sections of a document, or use the next traversal to suggest additional related terms for a document as its author is writing it.

%% file: future_work.tex
\section{Future Work}
While we have designed the SKG model, created and open sourced a reference implementation, and tested several use cases, there are many extension points worthy of future research and exploration.

Our implementation of the SKG is able to both identify predefined nodes, as well as materialize new nodes and edges on the fly. While we have implemented two general-purpose scoring mechanisms for assigning weights to edges (popularity and relatedness) and have implemented another more specialized scoring mechanism described in the career pathing use case section, each edge scoring algorithm must be coded into the system today.
A future extension of our implementation is to allow users to specify functions as part of the graph query syntax such that they can apply arbitrarily complex scoring calculations without the need to write custom code.

Additionally, whereas today all documents matching a term or query are included in materialized nodes and edges (even if they are only tangentially related to the document in which they were found), we believe that by first scoring all documents matching each node (for example, using a tf-idf score) and only leveraging the top $n$ documents from each node when scoring, that we could further improve the system's ability to identify highly-related nodes and to filter out noisy edges. 

\textit{Semantic Search}: One of the key future use cases where we intend to apply the SKG model is for query interpretation and expansion. We have already shown that the SKG works quite well for identifying the most conceptually similar terms for a given term/phrase/entity within a domain. This can be used to automatically expand a search query to perform a conceptual search instead of an exact match search. For example, if someone searches for \textit{cdl}, then the query could be automatically expanded to something like \textit{cdl OR "truck driver" OR freight OR "commercial drivers license"}. One particularly interesting aspect of using the SKG for this task is that not only can it identify what each individual search term means, but it can actually identify which terms are semantically-related to the entire query. Thus, if someone searches for \textit{driver AND windows}, the SKG can return a different set of keyword expansions for the term \textit{driver} than if the user had searched for \textit{driver AND truck}. This deviates from traditional taxonomy approaches, which often rely on fixed meanings of each word, whereas the SKG can infuse nuance and contextual interpretation of search terms.

\textit{Search Engine Relevancy Algorithms}: There are also several options for leveraging the technique mentioned in the document summarization use case to calculate and index the significance of each term in each document and use that information as part of the search engine's relevancy ranking algorithm. Such a probabilistic relevancy ranking function could likely achieve measurable gains over more traditional models which only consider the number of occurrences of terms as opposed to their conceptual significance to the document.

\input{trending_topics}

\input{recommendation_systems}

\input{root_cause_analysis}

\input{abuse_detection}

%% file: trending_topics.tex
\textit{Trending topics (time-series)}: Another application of the SKG is the identification of trends over time. This could be conceptually described as doing anomaly detection where the foreground and background sets are time frames as opposed to categories or keywords. For example, if you were analyzing a news feed or stream of social media posts, you could specify a background query of "this month" and a foreground query of "this week" or "today" to see articles or categories which are occurring with a higher-than expected likelihood. This would allow you to identify trending topics, and is a useful additional use case for future exploration with the SKG model.

%% file: recommendation_systems.tex
\textit{Recommendation Systems}:
Most recommendation systems leverage behavioral-based collaborative filtering, which suffers from the cold-start problem (items which have not yet been reviewed by enough users will not be recommended). In order to overcome this, it is often helpful to also have a content-based recommendations approach. The SKG can be leveraged to identify the most significant features of a document (as previously described in the document-summarization use case), in order to use those to match other documents sharing those same features. The SKG can also be leveraged to better understand the users for which recommendations are being made by inspecting the known information about them from their previous interactions with the system as compared with other users. For example, if a user ran multiple searches within a search engine before, the SKG can be used to determine the intersection and overlap between those searches (a materialized node) and to traverse to the other nodes that are most related to the combined search history of the user. Further, the SKG could be used to predict interests of users based upon how their behavior compares to other users. The career pathing use case described previously is a good example of this, where we could inspect a job seekers employment history and current job searches to determine, based upon other job seekers' typical behavior, when the user is most likely to switch jobs, and to what kind of job he/she would be willing to switch. This information could then be used as a feature in the job recommendation algorithm, and this feature would also change to more heavily favor a progression in job seniority as time progresses. Depending upon the domain, there are numerous ways to leverage the SKG by leveraging its ability to materialize and score the nuanced relationships between arbitrary entities.

%% file: root_cause_analysis.tex
\textit{Root Cause Analysis}:
The SKG is a good candidate for future research as a root cause analysis tool. Many companies maintain ticketing systems or online help forums through which they receive questions, bug reports, and/or complaints. The SKG could be used, for example, to find posts by customers matching a specific criteria (i.e. look for nodes/terms like \textit{frustrated}, \textit{broken}, or \textit{refund}) and find out which other words or topics are most statistically related. If a company sold a software system, this would be an easy way to determine which parts of the system needed the most attention.

%% file: abuse_detection.tex
\textit{Abuse Detection}:
Given a system where a few users partake in abusive behavior (posting spam, programmatically crawling the website, etc.), you could index the behaviors of users, find some abusive users, and use them as your foreground set to find other users who exhibited statistically similar behavior. In the spam use case, you could also tag your original documents with \textit{spam} or \textit{not spam} and set \textit{spam} documents as your foreground set $D_{FG}$. Assuming your documents contain textual content, you could then identify nodes/terms more commonly found among spam postings, and use this detection as the basis of a spam classifier for new postings as they are received. This is one of many forms of anomaly detection, a category of use cases for which the SKG is particularly well suited for future research and applications.

%% file: conclusion.tex
\vspace{-5pt}
\section{Conclusion}
In this paper, we have introduced a novel kind of knowledge discovery model, which we are calling the Semantic Knowledge Graph. This system enables the automatic creation of a graph which encodes statistical relationships between all keywords, phrases and entities represented within free text and semi-structured documents, allowing those relationships to be traversed and scored based upon strength of the relationship within a specific domain. This auto-generated graph can then be queried in real time to discover the nuanced relationships between any combination of linguistic representations (keywords, phrases, etc.) or structured data (titles, categories, dates, numbers, etc.). Unlike traditional graph databases which perform either a depth-first search or a breadth-first search of all nodes, because the Semantic Knowledge Graph materializes edges between nodes and assigns their weights on the fly based upon either count of overlapping documents or relatedness of nodes within a corpus of documents, the Semantic Knowledge Graph can use these weights to only traverse the top scoring edges. This turns the graph traversal process into one index lookup and set intersection per level of depth of the traversal, making the Semantic Knowledge Graph highly efficient at traversing millions or even billions of nodes, as long as only the highest weighted nodes are collected at each level.

The Semantic Knowledge Graph has numerous applications, including automatically building ontologies, identification of trending topics over time, predictive analytics on time-series data, root-cause analysis surfacing concepts related to failure scenarios from free text, data cleansing, document summarization, semantic search interpretation and expansion of queries, recommendation systems, and numerous other forms of anomaly detection. The main contribution of this paper is the introduction of the the Semantic Knowledge Graph, a novel and compact new graph model that can dynamically materialize and score the relationships between any entities represented within a corpus of documents.

%% file: acknowledgment.tex
\section*{Acknowledgment}
The authors would like to thank CareerBuilder for their sponsorship of this research and development. In particular, the authors thank Daniel Crouch, David Bernal, Lamar Payson, and Jacob Maggio, who also contributed their ideas and code reviews during the development of the semantic knowledge graph, as well as Colin Field, Matt McNair, Eric Presley, and Abdel Tefridj for their support of the development and open sourcing of the referenced implementation.